\documentclass[11pt]{article}
\usepackage{graphicx}
\DeclareGraphicsExtensions{.pdf,.png,.jpg}
\usepackage{amsmath}       
\usepackage{amsfonts}      
\usepackage{hyperref}      
\usepackage{geometry}      
\usepackage{fancyhdr}      
\usepackage{lineno}        
\usepackage{authblk}       
\usepackage{titlesec}      
\usepackage{cite}          
\usepackage{tcolorbox}  
\usepackage{makecell}
\usepackage{textcomp}

\definecolor{color1}{rgb}{0.014, 0.494, 0.549}   
\definecolor{color2}{rgb}{1.0, 1.0, 0.98}        
\definecolor{fontcolor}{rgb}{0.949, 0.298, 0.153} 
\title{
\vspace{-3cm} 
\begin{center}
    \makebox[\textwidth]{%
        \includegraphics[width=1.2\textwidth]{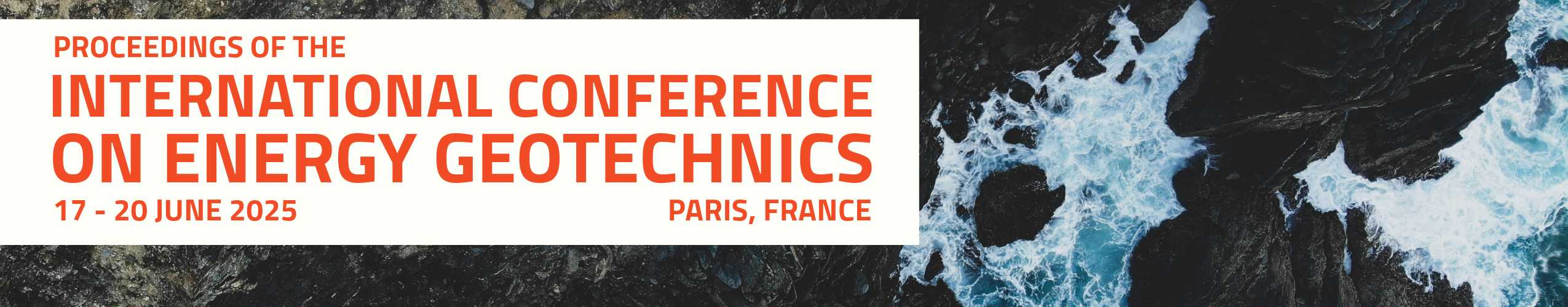} 
    } 
\end{center}
\noindent
\begin{tcolorbox}[colframe=white, colback=white, sharp corners, width=1.2\textwidth, boxrule=0mm, left=0mm, right=0mm, top=0mm, bottom=0mm]
\raggedright
\textnormal{\fontsize{9}{12}\selectfont © Authors: All rights reserved, 2025}  
\end{tcolorbox}
\vspace{0.5em} 
\centering 
Mineral Type Impact on Thermal Conductivity of Biocement and Biocemented Sand}  % Please insert your abstract title here

\author[1,*]{Shadi Zeinali} % Please adapt the authors and add if necessary
\author[1]{Zarghaam Heidar Rizvi} % The corresponding author should be indicated with an asterisk
\author[1]{Frank Wuttke}

\affil[1]{\fontsize{10}{12}\selectfont \itshape Kiel University, Ludewig-Meyn-Str.10, 24118, Kiel, Germany}
\affil[*]{\fontsize{10}{12}\selectfont \itshape Corresponding author: shadi.zeinali@ifg.uni-kiel.de}

\date{}  % please leave the date empty

\begin{document}
\maketitle
\thispagestyle{fancy}  

% adapt the contents of your mansucript in the following:

\section*{Summary}
This study experimentally investigates the influence of $CaCO_3$ polymorphs on the thermal conductivity (TC) of biocement and biocemented sand. Calcite-rich biocements consistently showed higher TC than vaterite-dominated ones, regardless of density or saturation. Vaterite-rich biocement, resulting from rapid precipitation and elevated organic content, yields inherently lower TC. This study shed light on the limited TC improvement observed in vaterite-rich biocemented sand.
\section{Introduction}
The global energy transition requires advanced materials to enhance the efficiency and sustainability of geothermal systems, a key renewable energy technology\cite{ref1}.
Microbially induced calcium carbonate precipitation (MICP) offers a promising method for improving soil thermal conductivity (TC) by forming $CaCO_3$ “thermal bridges” that increase particle contact and heat transfer pathways. Calcite, aragonite, and vaterite are the principal $CaCO_3$ polymorphs, with calcite and aragonite prevalent in nature, and vaterite the least stable\cite{ref2}. The possibility of precipitation and stabilization of vaterite within soil matrix has been shown in various studies \cite{ref2}-\cite{ref4}. Controlling or preserving its morphology is challenging; however, under right conditions it has the potential to recrystallize into calcite\cite{ref2}.
Recent studies have demonstrated that MICP can increase the TC of dry sands by up to 250–330\% at moderate calcite contents (6.7–8\%), although the scale of this effect diminishes near saturation. TC improvements also correlate with both calcification content and degree of saturation, while solutions composition, soil gradation, void ratio, fines content, and organics impacts its TC\cite{ref5}.
Existing research has primarily focused on the role of calcite in enhancing soil TC, leaving the influence of vaterite largely unexplored. Given that vaterite can precipitate in significant quantities under common MICP conditions\cite{ref4},\cite{ref6}, its impact on TC of biocement used in soil needs investigation. This study examines, for the first time, the specific impact of vaterite versus calcite polymorphs on the TC of biocement, and utilizes these findings to analyze the impact of vaterite-rich biocemented sand in our research\cite{ref7}. To address this literature gap, 5 biocement columns from different Bacterial suspension (BS) and Calcification solution (CS) constituents at different conditions were prepared, and tested in dried and saturated states by performing transient and steady-state TC tests on in-media precipitated biocement columns. 

\section{Materials and Methods}
%\subsection{Bacterial and Calcification solutions, and Sample preparation}
\textit{Sporosarcina pasteurii} (DSM33) was cultivated in Medium220 (German Collection of Microorganisms) with 20g/L (333mM) urea to prepare the bacterial medium (BM). The autoclaved medium was inoculated with 5\%(v/v) cell culture, incubated at 30°C for 72h, and stored at 4°C prior to use. Ammonium concentration, measured by autoanalyzer after incubation, indicated that 87\% of urea was hydrolyzed due to both bacterial ureolysis and autoclaving. The BM used daily had an average $OD_{600}$ of $1.10\pm0.29$ and urease activity of $1.9\pm0.58$ mM/min. Conditions and compositions of the bacterial suspension (BS) and calcification solution (CS) are detailed in Table~\ref{tab1}.

Biocementation experiments (Bioc2) targeted vaterite precipitation in sand columns \cite{ref7}. In each injection cycle, one quarter of the solution remained, and three quarters were replaced after 24h with fresh BS and CS at the same concentrations, assuming complete resource consumption. The process was repeated daily for seven cycles in a closed, magnetically stirred tank, monitoring pH and electrical conductivity (EC) in all steps. On day 7, precipitates were collected, oven-dried at 60°C to constant weight, powdered, remolded into columns. To monitor recrystallization during dry–saturation cycles, the procedure was repeated, with XRD subsamples collected after each drying step and polymorph proportions quantified by Rietveld analysis. Bioc0 followed the Bioc2 protocol but was stored at 4°C in a closed container for 1.5 years, simulating extended reaction at low temperature.  Bioc1, identical to Bioc2 in composition, but processed over two days to assess reduced reaction time. For Bioc3, the BM was filtered (0.2\textmu m PES membrane,$<$0.25bar vacuum), and the bacterial cells were resuspended in 0.9\% NaCl saline. Organic content was determined by loss-on-ignition (LOI, $\sim$10g at 550°C for 2h). TC was measured with a transient needle (TR-1, Decagon Devices) in dry and saturated states (24h water soak); selected samples were also tested by steady-state method to evaluate the potential impact of measurement methodology on TC values. 
    \begin{table}[ht]
    \centering
    \caption{Condition and compositions of BS and CS}
    \label{tab1}
    \begin{tabular}{cccc}
    \hline
    sample & condition & BS & CS \\
    \hline
    Bioc0 & long storage time at low T & BM & 85.5mM $CaCl_2$, pH=10 \\
    Bioc1 & short reaction time & BM & 85.5mM $CaCl_2$, pH=10\\
    Bioc2 & aimed high vaterite & BM & 85.5mM $CaCl_2$, pH=10\\
    Bioc3 & aimed high calcite & \makecell[l]{resuspended bacterial\\cells of BM in NaCl} & \makecell[l]{85.5mM $CaCl_2$\\+ 83.3mM urea}, pH=10 \\
    Bioc4 & aimed to increase calcite & BM & \makecell[l]{85.5mM $CaCl_2$\\+ 83.3mM urea}, pH=10 \\
    \hline
    \end{tabular}
    \end{table}   

\section{Results and discussion}
\begin{figure}[t]
\centering
\includegraphics[width=1\textwidth]{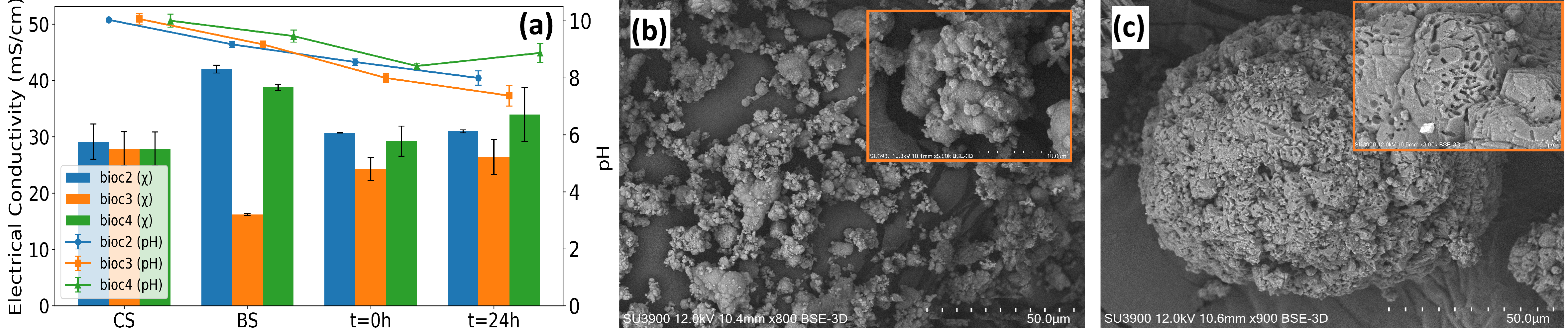} % Replace with your image file
\caption{(a) EC and pH evolution; SEM image of (b) Bioc2-V3, (c) Bioc3-V1 with 50\textmu m scale bar}
\label{fig1}
\end{figure}

\begin{figure}[t]
\centering
\includegraphics[width=1\textwidth]{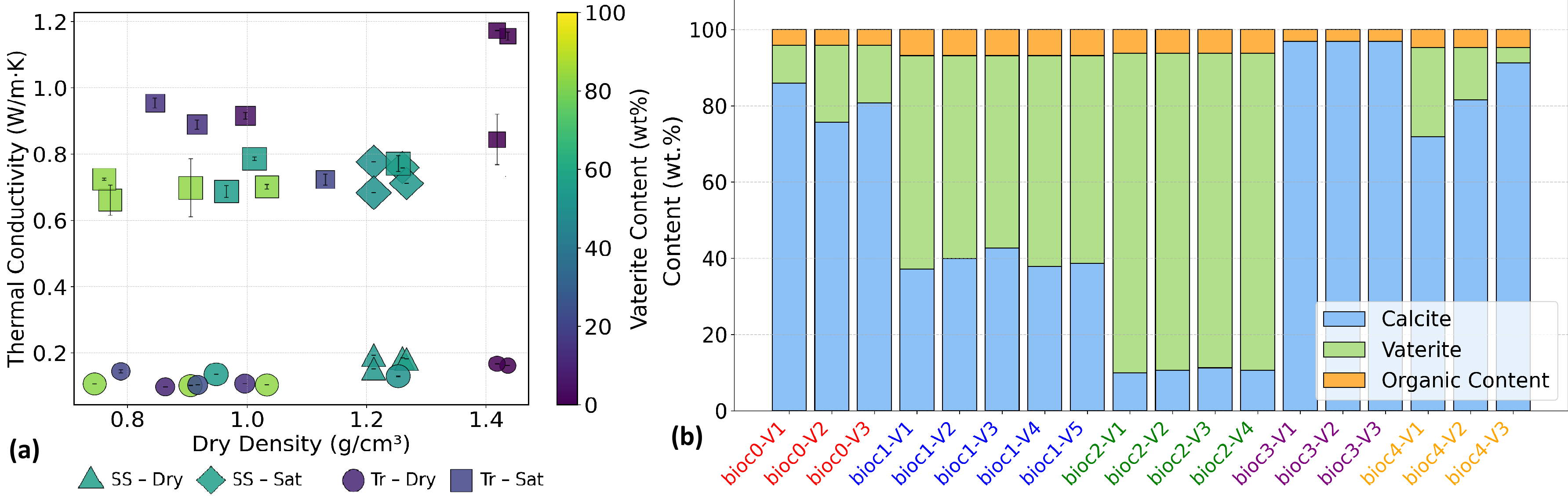} % Replace with your image file
\caption{(a) TC of biocement columns per dry density versus vaterite content, size of the markers is relative to organic content of the samples; (b) XRD, LOI composition of biocement samples.}
\label{fig2}
\end{figure}

Fig.~\ref{fig1}(a) shows EC and pH evolution during reaction, while SEM images (Fig.~\ref{fig2}(b,c)) contrast the microstructures of vaterite- and calcite-rich samples. Bioc2, with rapid EC decline and pH rise, formed predominantly vaterite with higher organic content, while Bioc3 and to a lesser extent Bioc4 showed more gradual changes, yielding calcite-rich products with lower organic matter. These solution chemistries determine precipitation kinetics, polymorph outcome, and organic incorporation, thereby explaining differences in microstructure and TC.
Combined SEM, XRD, and LOI analyses reveal that prolonged low-temperature storage converts vaterite to calcite (bioc2 to bioc0). Recrystallization during 24h saturation–drying cycles was greatest in bioc4, moderate in bioc0 and bioc1, and least in bioc2, likely due to its high initial supersaturation and organic content, which aligns with previous studies\cite{ref8}. Rapid precipitation in Bioc2 produced fine, grape-like spherical vaterite, whereas Bioc3 developed larger rhombic calcite crystals. Shorter reaction time (Bioc1) compared to Bioc2 resulted in a ~60:40 vaterite-to-calcite ratio, yielding intermediate density and TC in both dry and saturated states. Current study, limited to specific drying–saturation cycles and storage conditions, demonstrates clear polymorph influence on TC. Additional samples and broader environmental conditions are required and currently under preparation to validate these trends for engineering application.

TC assessment of biocement columns depicted in Fig.~\ref{fig2}(a), shows a clear dependence on their polymorph composition. Calcite-rich samples (notably Bioc3 and Bioc4) consistently exhibited higher TC than vaterite-dominated Bioc2 under both dry and saturated conditions, with differences amplified at higher dry densities. Quantitatively, in dry-state, Bioc3 (highest calcite content, $\sim$1.4g/cm$^3$) reached a TC of $\sim$0.17W/mK, about 1.6 times of vaterite-dominated Bioc2 ($\sim$0.1W/mK at $\sim$0.75-1.0g/cm$^3$). Under saturated condition, all samples experienced the expected rise in TC due to water filling pores and water’s higher TC relative to air; however, the calcite-rich samples still outperformed the vaterite-rich ones. In each dry density, the more vaterite content is the less TC is achieved. Increasing the calcite fraction from roughly 10\% to 90\% resulted in $\sim$30\% increase in TC (Bioc2 compared to Bioc4 samples in same range of density $\sim$0.9-1g/cm$^3$). Higher organic content, particularly in the saturated state, was associated with reduced TC. SEM images confirmed finer, spherical vaterite crystals in Bioc2 versus larger, rhombic calcite in Bioc3. It can be concluded that the polymorph type significantly influences thermal behavior. TC measurements using both transient (Tr) and steady-state (SS) methods were in good agreement. These findings indicate that vaterite presence, high organic content, and lower density all contribute to reduced TC in biocement. These findings suggest that the comparatively modest increase in TC observed in biocemented sand samples with up to 10\%(g/g) calcification in the authors’ previous study \cite{ref7}, relative to calcite-rich biocemented sands reported in the literature, can be attributed -at least in part- to the predominance of vaterite, as confirmed by XRD (exceeding 95\% of the precipitate). Given vaterite's intrinsic lower stability, higher solubility, distinct lattice structure, and lower density compared to calcite \cite{ref2}, the present findings indicate that its dominance fundamentally restricts achievable improvements in TC.

\section{Conclusion}
Vaterite-rich biocement exhibits markedly lower TC compared to calcite-rich variants, due to differences in crystal structure, density, and organic incorporation. This mineralogical effect likely underlies the modest TC increases observed in vaterite-rich biocemented sands. To enable predictive engineering design, further studies on biocement and biocemented soils with controlled polymorph mixtures are required. Future research with controlled polymorph ratios and diverse field-relevant conditions will strengthen predictive models for geothermal engineering design.

% References. Alternatively, use \bibliographystyle{IEEEtran}

\end{document}